\documentclass[12pt,notoc]{JHEP}
\usepackage{amssymb}
\title{Mass hierarchy and localization of gravity in extra time
\thanks{Supported by the Academy of Finland under the Project No. 163394}}
\author{Masud Chaichian and Archil B. Kobakhidze \thanks{On leave of absence from
Andronikashvili Institute of Physics, Tbilisi, Georgia.} \\ High
Energy Physics Division, Department of Physics, University of
Helsinki and \\ Helsinki Institute of Physics, P.O. Box 9,
FIN-00014 Helsinki, Finland \\ Email: {\rm
Masud.Chaichian@helsinki.fi, Archil.Kobakhidze@helsinki.fi}}
\abstract{We consider Randall-Sundrum model with localized gravity,
replacing the extra compact space-like dimension by a time-like one.
In this way the solution to the hierarchy problem can be
reconciled with a correct cosmological expansion of the visible
universe, just as a trivial result of the sign flip of cosmological
constants in the bulk and on the 3-branes relative to the case of
extra space-like dimension. Some phenomenological aspects of the proposed scenario
related to the tachyonic nature of Kaluza-Klein states of graviton are also
discussed.}
\preprint{HIP-2000-16/TH}
\begin{document}
During the few past years it has been realized that the
fundamental scales of physics can be altered in the presence of
extra dimensions \cite{1,2,3}. What is exceptionally exciting is
that the fundamental Planck scale \cite{2} or/and fundamental GUT
scale \cite{3} can be lowered to potentially accessible energies
in the multi-TeV range. In addition to the well known explanations
within the supersymetric models or the models with dynamical
symmetry breaking, these observations offer a qualitatively new
explanation of the observed hierarchy between the electroweak
scale and high energy scales. Despite the important differences
between these two scenarios, both similarly utilize $\delta$ extra
compact dimensions with large compactification radii $r_{n}$
($n=1,...,\delta$) in the factorizable, $M^{4}\times N^{\delta}$, 
($4+\delta$)-dimensional spacetime and thus the apparent weakness of
gravity in the visible four-dimensional world ($M^{4}$) is
explained due to the large volume $V_{N^{\delta}}$ of the 
extra-dimensional submanifold $N^{\delta}$ (for an earlier proposal of  
large extra dimensions, see \cite{an})\footnote{It has been recently 
proposed \cite{4} that the vacuum expectation value of the
electroweak Higgs boson can be exponentially suppressed due to the
renormalization effects in higher dimensional theories, thus
explaining Planck/weak scale hierarchy without need of
hierarchically large extra dimensions. For another approach to
solve the hierarchy problem within the higher dimensional gauge
theories see \cite{5}.}:
\begin{equation}
M_{Pl}^{2}= M_{\ast}^{\delta + 2}V_{N^{\delta}},
\label{1}
\end{equation}
where $M_{\ast}$ is the fundamental high-dimensional scale and 
$M_{Pl}$ is the ordinary four-dimensional Planck scale.

More recently, a new scenario \cite{6} for generating Planck/weak
scale hierarchy has been proposed within the framework of
5-dimensional non-factorizable AdS$_5$ space-time with two 3-branes 
located at the $S^{1}/Z_{2}$ orbifold fixed points of the fifth
compact dimension. Now the weakness of gravity in the visible
world 3-brane is explained without recourse to large extra
dimensions, but rather as a result of gravity localization on the
hidden 3-brane. Gravity localization in such scenario occurs
because the five-dimensional Einstein's equations admit the
solution for the space-time metric with a scale factor ("warp
factor") which is a falling exponential function of the distance
along the extra dimension y perpendicular to the
branes\footnote{Earlier, it was suggested in \cite{7} that
gravitational interaction between particles on a brane in
uncompactified five-dimensional space could have a correct
four-dimensional Newtonian behaviour, provided that the corresponding
contributions to Newton's law from the bulk cosmological
constant and from the brane tension cancel each other. For some
previous related works, see \cite{8}.}:
\begin{equation}
ds^{2}=e^{-2k|y|}dx^{2}_{1+3}+dy^{2},
\label{2}
\end{equation}
when the bulk cosmological constant $\Lambda$ ($\Lambda < 0$) and
the tensions $T_{vis}$ and $T_{hid}$ of the visible and hidden
branes respectively are related according to\footnote{Actually
this relation is nothing but the condition for the vanishing of the
four-dimensional effective cosmological constant.}:
\begin{equation}
T_{hid}=-T_{vis}=6M^{3}_{\ast}k,~~~
k=\sqrt{-\frac{\Lambda}{6M^{3}_{\ast}}}. \label{3}
\end{equation}
Thus, graviton is essentially localized on the hidden brane with
positive tension ($T_{hid} > 0$) which is located at $y=0$ fixed
point of the $S^{1}/Z_{2}$ orbifold, while the Standard Model
particles are assumed to be restricted on the visible brane with
negative tension ($T_{vis} < 0$) which is located at $y=\pi r_{c}$
($r_{c}$ is the size of extra dimension) orbifold fixed point. So,
a hierarchically small scale factor generated for the metric on
the visible brane gives an exponential hierarchy between the mass
scales of the visible brane and the fundamental mass scale
$M_{\ast}$, after one appropriately rescales the fields on the
visible brane. In fact, assuming $M_{\ast}\sim M_{Pl}$, TeV-sized
electroweak scale can be generated on the visible brane by requiring
$r_{c}\cdot M_{\ast}\simeq 12$. Various modifications and
generalizations as well as interesting phenomenological and
cosmological aspects of this scenario are intensively discussed in
the literature [9-16].

Soon after Ref. \cite{6} appeared it was pointed out in \cite{9}
that having a negative tension visible brane would be problematic
from the cosmological point of view, since Friedmann's equation
governing the expansion of the visible universe appears with wrong
sign. In fact, Einstein's equations posses another solution
\cite{7,10} which can be obtained from (\ref{2}) by changing the sign
of the $k$ parameter:
\begin{equation}
k\rightarrow -k.
\label{4}
\end{equation}
Since the transformation (\ref{4}) is not a symmetry of the theory
(unless simultaneously accompanied by the shift $y\rightarrow
y+\pi r_{c}$) the solution with $k<0$ is physically distinct from
the solution with $k>0$ and, as evident from (\ref{3}),
replacement (\ref{4}) exchanges the signs of brane tensions, so
that the visible brane at $y=\pi r_{c}$ becomes now the one with
positive tension. However, while the solution with $k<0$
\cite{7,10} is consistent with a Friedmann-like expanding universe,
the generation of Plank/weak scale hierarchy becomes now
impossible. To reconcile the cosmological expansion with the
solution of the Plank/weak scale hierarchy problem, more complex
constructions have been subsequently considered
\cite{11,12}\footnote{It was realized later that the solution to
the problem of correct cosmological expansion of the visible brane
can be linked to the problem of stabilization of extra space
\cite{12}. See \cite{13} for the stabilization mechanisms.}.

In this letter we would like to suggest that the problems of mass
hierarchy and cosmological expansion can be simultaneously solved
just in the frame of the original proposal of Ref. \cite{6} by simply
replacing the extra space-like dimension by a time-like
one\footnote{Extra time-like dimensions have been a subject of
interest for some time \cite{17} and have been revived recently
within the various versions of string and M-theory \cite{18,19}
and the so-called Two-Time Physics \cite{20,21}.}. Our solution
arises from a rather simple observation: The replacement
of the space-like dimension by a time-like one, $y\rightarrow
\tau$, i.e. the change of the signature from $(- + + + +)$ to $(-
+ + + -)$ leaves Einstein's equations unchanged if it is
simultaneously accompanied by the change of the sign of bulk
cosmological constant $\Lambda$ and the brane tensions $T_{hid}$ and
$T_{vis}$:
\begin{eqnarray}
(- + + + +)\rightarrow (- + + + -) \nonumber \\ \Lambda
\rightarrow -\Lambda,~~ T_{hid}\rightarrow -T_{hid},~~
T_{vis}\rightarrow -T_{vis}. \label{5}
\end{eqnarray}
Thus, in our scenario the AdS$_5$ space is replaced by the dS$_5$ one and 
the solution for the metric
\begin{equation}
  ds^{2}=e^{-2k|\tau|}dx^{2}_{1+3}-d\tau^{2}
\label{6}
\end{equation}
leads to the localization of gravity on the hidden brane with negative
tension stuck at the $\tau =0$ fixed point of the time-like
$S^{1}/Z_{2}$ orbifold, while the Standard Model particles are
placed on the positive tension brane at $\tau =\pi \tau_{c}$ with
\begin{equation}
-T_{hid}=T_{vis}=6M^{3}_{\ast}k,~~~
k=\sqrt{\frac{\Lambda}{6M^{3}_{\ast}}}. \label{7}
\end{equation}
The Planck/weak scale hierarchy is explained even with a small (in
Planck mass units) period of extra time (such as, $\tau_{c}\cdot
M_{Pl}\approx 12$), in full analogy with the case of space-like
extra dimension \cite{6}, while the positivity of the visible
brane tension ($T_{vis}>0$) ensures the correct Friedmann-like
expansion.

Despite the similarity of solutions with extra space-like
(\ref{2}) and extra time-like dimension (\ref{6}), the
phenomenological consequences of these two scenarios drastically
differ. As it is well known, typically theories with extra
time-like dimensions suffer from pathologies such as negative-norm 
states (ghosts) and tachyons\footnote{It was shown also that most
of theories with extra time have instantonic solutions \cite{22}
which may lead to the instability of flat space, but these 
solutions can be reinterpreted \cite{19} so that the question of
vacuum instability remains unclear.}. In fact the Kaluza-Klein
(KK) excitations in the case of compact extra time-like dimensions
would be seen by the four-dimensional observer as tachyonic 
states with imaginary masses quantized in units of 
$\frac{i}{\tau_{c}}$. The exchange of such KK states induces an
imaginary part in the effective low-energy potential between two
test "charges". This complexity was interpreted in \cite{23} as a
violation of causality and probability in the interaction of two
"charged" particles, so they can disappear into "nothing". If so,
from the experiments dedicated to look for proton or double
$\beta$ decays one can put rather severe bounds on the size of
extra time-like dimension, $\tau_{c}\lesssim 10\cdot M^{-1}_{Pl}$
\cite{23}, in the case of appearance of tachyonic KK states of
photon or gluons. Recently, however, phenomenological constraints
on extra time-like dimensions have been revisited in the framework
of brane world (with factorizable spacetime), where the only
particle feeling the extra time(s) is the graviton \cite{24}. It was 
argued there, that the induced imaginary part of the gravitational
potential can be reinterpreted as an artifact of the fictitious
decay into the unphysical negative energy tachyons and thus the
size of extra time-like dimensions can be as large as
$\tau_{c}\sim 1mm$! Since the graviton KK spectrum is quite
different in the case with non-factorizable geometry considered
here, we shall discuss now the phenomenology of the extra time in
more details.

Let us first determine the mass spectrum of the graviton KK modes
in the effective four-dimensional theory. The starting point is
the five-dimensional Einstein equations
\begin{eqnarray}\label{8}
\sqrt{G}(R_{MN}-\frac{1}{2}G_{MN}R)= \nonumber \\
-\frac{1}{M^{3}_{\ast}}[\Lambda \sqrt{G}G_{MN}+
T_{vis}\sqrt{-g_{vis}}g^{vis}_{\mu\nu}\delta^{\mu}_{M}\delta^{\nu}_{N}\delta(\tau
-\pi \tau_{c})+
T_{hid}\sqrt{-g_{hid}}g^{hid}_{\mu\nu}\delta^{\mu}_{M}\delta^{\nu}_{N}\delta(\tau)],
\end{eqnarray}
where $G_{MN}$ ($M,N=\mu , \tau$) is the five-dimensional metric
and $g^{vis}_{\mu\nu}=G_{\mu\nu}(x^{\mu}, \tau =\pi)$ and
$g^{hid}_{\mu\nu}=G_{\mu\nu}(x^{\mu}, \tau =0)$ are
four-dimensional metrics on the visible and hidden branes,
respectively. It is easily verified that the actual solution to
eq.(\ref{8}) is given by (\ref{6}) with condition (\ref{7})
satisfied. Now let us look at the linear perturbations about this
solution which can be parametrized by replacing $\eta_{\mu\nu}$
with $\eta_{\mu\nu}+h_{\mu\nu}(x,\tau)$. Upon compactification, the 
graviton field $h_{\mu\nu}(x,\tau)$ can be expanded into a KK tower 
as:
\begin{equation}\label{9}
h_{\mu\nu}(x,\tau)=\sum^{\infty}_{n=0}h^{(n)}_{\mu\nu}(x)\psi^{(n)}(\tau),
\end{equation}
where $h^{(n)}_{\mu\nu}(x)$ are KK modes of the graviton on the
background of Minkowski space on the 3-brane. In the transverse
traceless gauge ($\partial_{\mu}h^{\mu\nu}=h^{\mu}_{\mu}=0$) the
equation of motion for $h^{(n)}_{\mu\nu}(x)$ is given by:
\begin{equation}\label{10}
(\eta^{\mu\nu}\partial_{\mu}\partial_{\nu}+m^{2}_{n})h^{(n)}_{\alpha\beta}(x)=0.
\end{equation}
Note that in contrast to the case of extra space-like dimension
the sign of $m^{2}_{n}$ in (\ref{10}) is flipped and thus this
equation of motion describes graviton KK states with imaginary
masses, i.e. tachyonic gravitons \cite{24}. Imposing
orthonormality condition for $\psi^{(n)}(\tau)$, 
\begin{equation}\label{11}
\int^{\pi\tau_{c}}_{-\pi\tau_{c}}d\tau
e^{-2k|\tau|}\psi^{(m)}\psi^{(n)}=\delta_{mn},
\end{equation}
Einstein's equations (\ref{8}) in conjunction with the above
equation of motion (\ref{10}) give the following differential
equation for $\psi^{(n)}(\tau)$:
\begin{equation}\label{12}
\frac{d}{d\tau}(e^{-4k|\tau|}\frac{d\psi^{(n)}}{d\tau})=-m^{2}_{n}e^{-2k|\tau|}\psi^{(n)}.
\end{equation}
This is just the same equation as in the case of extra space-like
dimension \cite{10}. The solution to the eq.(\ref{12}) is
expressed by the Bessel functions of order two:
\begin{equation}\label{13}
\psi^{(n)}(\tau)=\frac{e^{2k|\tau|}}{N_{n}}[J_{2}(z_{n})+A_{n}Y_{2}(z_{n})],
\end{equation}
where $z_{n}(\tau)=\frac{m_{n}}{k}e^{k|\tau|}$, $N_{n}$ is the
normalization factor and $A_{n}$ is a constant. The
boundary conditions $\frac{d}{d\tau}\psi^{(n)}(\tau)|_{\tau
=0,\pi}=0$ lead to the following equations:
\begin{equation}\label{14}
A_{n}=-\frac{J_{1}(z_{n}(0))}{Y_{1}(z_{n}(0))},
\end{equation}
\begin{equation}\label{15}
A_{n}=-\frac{J_{1}(z_{n}(\pi))}{Y_{1}(z_{n}(\pi))},
\end{equation}
through which one can determine $A_{n}$ and $m_{n}$. In fact,
working in the limit $z_{n}(0)\ll 1$, one finds $A_{n}\sim
z_{n}(0)^{2} \approx 0$ and $J_{1}(z_{n}(\pi))\approx 0$. Thus, the
masses of the graviton KK modes given by
$m_{n}=kz_{n}(\pi)e^{-k\tau_{c}\pi}$, are essentially determined
through the roots of $J_{1}(z_{n}(\pi))$ and generally are not
equally spaced, but in the limit $z_{n}(\pi)\gg 1$,
$J_{1}(z_{n}(\pi))$ is approximated by $\sqrt{2/\pi
z_{n}(\pi)}\cos(3\pi /4 - z_{n}(\pi))$ and thus,
\begin{equation}\label{16}
\Delta m=m_{n+1}-m_{n}\approx \pi k e^{-k\tau_{c}\pi}.
\end{equation}
Finally, from (\ref{11}) one finds the normalization:
\begin{equation}\label{17}
N_{n}\approx \frac{e^{\pi
k\tau_{c}}}{\sqrt{k}}|J_{2}(z_{n}(\pi))|
~~~_{\overrightarrow{~~_{z_{n}(\pi)\rightarrow \infty }~~}~~~}
\frac{e^{\pi
k\tau_{c}}}{\sqrt{k}}\sqrt{\frac{2}{\pi z_{n}(\pi)}}.
\end{equation}
The zero mode wave function can be easily obtained from the general
solution (\ref{13}) by the limiting procedure $m_{n}\rightarrow
0$:
\begin{equation}\label{18}
\psi^{(0)}=\sqrt{\frac{k}{1-e^{-2\pi k\tau_{c}}}}.
\end{equation}

Following Refs. \cite{9,14}, we have also found a non-static
(cosmological) solution and calculated the four-dimensional Hubble
constant
\begin{equation}\label{19}
H^{2}=
\frac{(T_{hid}+\varrho_{hid})^{2}}{36M^{6}_{\ast}}-\frac{\Lambda}{6M^{3}_{\ast}}=
\frac{(T_{vis}+e^{-2\pi
k\tau_{c}}\varrho_{vis})^{2}}{36M^{6}_{\ast}}-\frac{\Lambda}{6M^{3}_{\ast}},
\end{equation}
where $\varrho_{hid}$ and $\varrho_{vis}$ ($\varrho_{vis}=-e^{2\pi
k\tau_{c}}\varrho_{hid}$) are matter energy densities on the
hidden and visible branes, respectively. Taking into account
(\ref{7}) and $e^{-4k\pi \tau_{c}}\varrho_{hid}/M^{4}_{\ast}\ll 1$, one
obtains the desired form for the Hubble constant on the visible
brane, $H^2\approx T_{vis}e^{-2\pi k\tau_{c}}\varrho_{vis}/(18M^{6}_{\ast})$,
since now $T_{vis}>0$.

Having determined the KK spectrum of the effective four-dimensional,
theory we are ready now to discuss the possible
influences of extra time on the ordinary four-dimensional physics.
The corrections appeared due to the graviton KK modes exchange to
the gravitational potential of the two test point masses $M_{2}$
and $M_{1}$ placed at the points $(x=0, \tau =\tau_{c}\pi)$ and $(|x|=r, \tau
=\tau_{c}\pi)$ of the visible 3-brane, can be expressed as:
\begin{eqnarray}\label{20}
V(r)=\sum^{\infty}_{n=0}G^{(5)}_{N}\frac{M_{1}M_{2}}{r}|\psi^{(n)}(z_{n}(\pi))|^{2}e^{-im_{n}r}
=G_{N}\frac{M_{1}M_{2}}{r}+\delta V(r), \nonumber \\ \delta
V(r)=\sum^{\infty}_{n=1}G^{(5)}_{N}\frac{M_{1}M_{2}}{r}|\psi^{(n)}(z_{n}(\pi))|^{2}e^{-im_{n}r},
\end{eqnarray}
where the five-dimensional Newton constant
$G^{(5)}_{N}=1/M^{3}_{\ast}$ is related to the ordinary
four-dimensional one $G_{N}=1/M^{2}_{Pl}$ as
\begin{equation}\label{21}
G_{N}=G^{(5)}_{N}k(1-e^{-2\pi k\tau_{c}})^{-1}.
\end{equation}
Thus, an imaginary part is induced in the Newton's potential
(\ref{20}) as a result of tachyonic nature of the graviton KK
modes \cite{24}. Typically, such complex contributions to the
energy are associated with an instability of the system. Let us
consider, for example, two neutrons inside a nucleus. Taking the
wave function to be \cite{23}
\begin{equation}\label{22}
\psi
(r)=\frac{m^{3/2}_{\pi}}{\sqrt{\pi}}e^{-m_{\pi}r},
\end{equation}
(here $m_{\pi}$ is the pion mass) we can calculate the gravitational
energy of the system corresponding to (\ref{20}):
\begin{equation}\label{23}
E=\langle \psi|V(r)|\psi \rangle,
\end{equation}
the imaginary part of which can be identified with a decay width of a
neutron into "nothing" \cite{23}:
\begin{equation}\label{24}
\Gamma =
\frac{16\zeta(3)}{\pi^{3}k^{3}}m^{2}_{N}m^{4}_{\pi}G_{N}e^{5\pi
k\tau_{c}},
\end{equation}
where $m_{N}$ is the neutron mass, $\zeta(3)\approx 1.2$ is a
Riemann's function value. In deriving (\ref{24}) we use planar wave
approximation to (\ref{13}) with almost equidistantly distributed
masses (\ref{16}). Now taking, e.g. $k\tau_{c}\approx 12$ (and $k\approx
M_{Pl}$), as it is desired for the solution of the hierarchy
problem, we get the life-time for the disappearance of a neutron
from a nucleus:
\begin{equation}\label{25}
\tau_{N}\approx 10^{-7} {\rm s.}
\end{equation}
Needless to say, the value (\ref{25}) is too low to be
consistent with present observations. The experimental lower
bound on the partial life-time of the decay $n\rightarrow
\nu\nu\bar\nu$ is 40 orders of magnitude larger than (\ref{25}).
Thus, the extra time-like dimension even with a small size
($\tau_{c}\simeq O(10)M^{-1}_{Pl}$) which would be
consistent with experiments in the case of factorizable geometry,
sharply contradicts the current experimental observations on matter stability in the case of
non-factorizable geometry considered here.

Let us note that the violation of probability in processes like
the ones considered above just follows from the expected violation
of causality in theories with extra time-like dimensions. However,
violation of causality is not an indisputable consequence of the
existence of extra time-like dimensions and deserves further
study. Indeed, violation of causality can be viewed as a result of
propagation of tachyonic KK modes of graviton with negative
energies backward in ordinary time. Clearly, to get a consistent
theory of tachyons, one could somehow remove the negative energy
tachyonic states from the physical spectrum (for some earlier
attempts, see \cite{25}). We do not aim here to go into the
details of tachyonic physics, but would like to simply note that
it seems quite reasonable that any solution to the problem of
negative energy tachyonic states would automatically solve the
problem of violation of causality. In that case, the
above-mentioned phenomenological inconsistencies can be
disregarded.

To conclude, we consider some cosmological and phenomenological aspects of the Randall-Sundrum model
with extra time-like dimension. We show that the introduction of the extra time-like dimension, instead of the
extra space-like one previously proposed, helps to reconcile the solution to the hierarchy problem with the correct
cosmological expansion of the visible universe. At the same time, we are faced with the problem of matter
instability related to a possible violation of causality and probability which is typical for the
theories with extra time-like dimensions, although, as stressed above, a clear-cut conclusion is by far less
obvious and the problem deserves further study \cite{26}.

\acknowledgments

We thank M. Gogberashvili and A. Schwimmer for useful discussions and comments.

\end{document}